# $Q$ factor in numerical simulations of DPSK with optical delay demodulation

Xing Wei, Xiang Liu, and Chris Xu

*Abstract*—A simple model is used to estimate the $Q$ factor in numerical simulations of differential phase shift keying (DPSK) with optical delay demodulation and balanced detection. It is found that an alternative definition of $Q$ is needed for DPSK in order to have a more accurate prediction of the bit error ratio (BER).

## I. INTRODUCTION

DIFFERENTIAL PHASE SHIFT KEYING (DPSK) was studied in the early days of optical fiber communications [1-4]. Recently, direct-detection (no local oscillator) DPSK has attracted much attention for applications in high bit-rate wavelength-division-multiplexed (WDM) systems, and studies were carried out both numerically [5-7] and experimentally [8-10]. Using DPSK, a transmission distance of 4,000 km at 42.7 Gb/s bit-rate has been demonstrated [10].

It is well-known that in a linear channel DPSK has a receiver sensitivity advantage over on-off keying (OOK) by approximately 3 dB with optical delay demodulation and balanced detection [11-14]. However, to our knowledge, such benefit of balanced detection has not been properly taken into account in DPSK numerical simulations published so far. For example, it was reported in [5] that "at the level of the practical optical-signal-to-noise (OSNR) regime, balanced detection gives only marginal improvement", which seems to contradict the results of earlier theoretical and experimental studies of DPSK.

We note that one challenge in numerical simulations is to provide a reliable estimate of the BER. To save the computation time, a typical simulation program uses only hundreds of bits, and therefore the BER is usually not counted directly but estimated by evaluating the statistical fluctuation in the received signal. In simulations of OOK, such fluctuation is often characterized by a $Q$ factor defined as

$$Q = \frac{|\mu_1 - \mu_0|}{\sigma_1 + \sigma_0}, \quad (1)$$

where $|\mu_1-\mu_0|$ denotes the separation between the intensity levels of "1" and "0", and $\sigma_1+\sigma_0$ is the sum of the standard deviations of the intensities around the levels of "1" and "0". Based on the Gaussian approximation for the noise distribution in the received signal, one can derive the relation between BER and $Q$, i.e.,

$$\text{BER} = \frac{1}{2}\text{erfc}\left(\frac{Q}{\sqrt{2}}\right) \approx \frac{1}{\sqrt{2\pi}Q}\exp\left(-\frac{Q^2}{2}\right). \quad (2)$$

For OOK, it is known that this method coincidentally gives a fairly good prediction of the BER, although the noise distribution in the intensity domain is not exactly Gaussian [11]. However, as will be explained in this Letter, direct use of (1) and (2) in simulations of DPSK may lead to wrong prediction of the BER even in the linear regime. This is not due to the patterning effect which has been discussed before for OOK [15], but due to the fundamentally non-Gaussian nature of the noise distribution in the output signal of the DPSK balanced receiver.

In Sec. II, we present a simplified model for a rigorous analysis of the BER in DPSK and show how the $Q$ factor definition (1) can be modified to predict the BER consistently. In Sec. III, we consider the effect of the nonlinear phase noise, which is the dominating nonlinear penalty in DPSK [16].

## II. DPSK IN A LINEAR CHANNEL

In a linear optical transmission system with optical amplifiers, the field of $N$ return-to-zero (RZ) pulses at the end of the transmission can be expressed as

$$F(t) = \left[\sum_{n=0}^{N-1} a_n u(t-nT) + z(t)\right] e^{-i\omega_c t} + \text{c.c.}, \quad (3)$$

where $\omega_c$ is the angular frequency of the optical carrier, $T$ is the bit period, $u(t-nT)$ is the envelope function of the RZ pulse in the $n$-th timeslot, $a_n$ is the (complex) amplitude of the $n$-th pulse, and $z(t)$ represents a classical additive Gaussian noise (the optical noise with a different polarization will be neglected for simplicity). For DPSK, the information is encoded in a relative phase change of the signal amplitude $a_n$ with respect to the previous symbol $a_{n-1}$. In this Letter, we choose $a_n = \pm 1$, a digital "0" will be represented by a phase change of $\pi$ or $a_n = -a_{n-1}$, and a digital "1" will be represented by no phase change or $a_n = a_{n-1}$.

To optimize the receiver performance, in front of the receiver we use a matched optical filter with an impulse

Xing Wei is with Lucent Technologies, Bell Laboratories, Murray Hill, NJ 07974 USA (e-mail: xingwei@lucent.com). Xiang Liu and Chris Xu are with Lucent Technologies, Bell Laboratories, Holmdel, NJ 07733 USA.

response function

$$h(t) = \frac{1}{\sqrt{E_b}} u(-t) e^{i\omega_c t} + \text{c.c.} \; , \tag{4}$$

where $E_b$ is the energy per bit

$$E_b = \int_{-T/2}^{T/2} u^*(t) u(t) dt \; . \tag{5}$$

The filtered signal is a convolution of $F(t)$ and $h(t)$. Near the center of the $n$-th timeslot, the filtered signal is

$$F_n(t) = f_n e^{-i\omega_c t} + \text{c.c.} \; , \tag{6}$$

$$f_n = a_n \sqrt{E_b} + z_n \; , \tag{7}$$

$$z_n = \int_{(n-1/2)T}^{(n+1/2)T} z(t) u^*(t - nT) dt \; . \tag{8}$$

The filtered noise amplitude $z_n$ consists of a real part $x_n$ and an imaginary part $y_n$, or

$$z_n = x_n + i y_n \; , \tag{9}$$

and $x_n$ and $y_n$ are independent zero-mean Gaussian-distributed quantities with the same variance

$$\langle x_n^2 \rangle = \langle y_n^2 \rangle = \sigma^2 \; . \tag{10}$$

It can be easily proven that $2\sigma^2$ is equal to the power spectral density $N_0$ of the unfiltered white noise $z(t)$ (single polarization), although this is not critical for this Letter.

The filtered DPSK signal can then be decoded with an optical delay interferometer as shown in Fig. 1. The optical output of the delay interferometer is either a constructive interference or a destructive interference depending on the relative phase between $f_n$ and $f_{n-1}$. The signals measured by the two photodiodes are

$$I_+ = \left| \frac{f_n + f_{n-1}}{2} \right|^2 , \tag{11}$$

$$I_- = \left| \frac{f_n - f_{n-1}}{2} \right|^2 . \tag{12}$$

A subtraction between $I_+$ and $I_-$ is then performed by a differential amplifier, and the balanced output is

$$I_{\text{bal}} = I_+ - I_- = \frac{f_n f_{n-1}^* + f_n^* f_{n-1}}{2} \; . \tag{13}$$

Using (7) and (9), we find

$$I_{\text{bal}} = \left( a_n \sqrt{E_b} + x_n \right)\left( a_{n-1} \sqrt{E_b} + x_{n-1} \right) + y_n y_{n-1} \; . \tag{14}$$

We note that the above expression represents the "inner product" of two vectors associated with the two complex quantities $f_n$ and $f_{n-1}$. Depending on the relative sign of $a_n$ and $a_{n-1}$, $I_{\text{bal}}$ is either around $E_b$ (for "1") or around $-E_b$ (for "0"), and the decision level is at zero. The BER is the probability for $I_{\text{bal}}$ to have a wrong sign. Using the Gaussian probability density function (PDF) of $x_n$ and $y_n$, one can calculate the BER analytically. The result of the calculation is

$$\text{BER} = \frac{1}{2} \exp\left( -\frac{E_b}{2\sigma^2} \right), \tag{15}$$

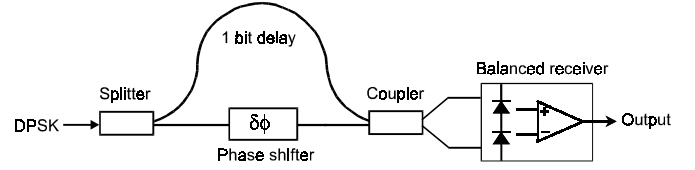

Fig. 1. Optical demodulation of DPSK with a delay interferometer and a balanced receiver.

which agrees with earlier results for an ideal DPSK system [1, 13]. We further note that (15) is only valid for DPSK with balanced detection. If only one output of the delay interferometer ($I_+$ or $I_-$) is used for direct detection, the performance of DPSK would degrade by approximately 3 dB and become equivalent to OOK.

Now we compare the analytical BER result (15) with what one would obtain by estimation using the $Q$ factor defined in (1). From (14), we find the balanced receiver output $I_{\text{bal}} \approx \pm E_b$, corresponding to $|\mu_1 - \mu_0| \approx 2 E_b$. Since $x_n$ and $x_{n-1}$ are independent, we find that the standard deviation of $I_{\text{bal}}$ is $\sigma_1 = \sigma_0 \approx \sqrt{2 E_b} \sigma$ in the small noise limit. Using these values in (1) then yields

$$Q \approx \frac{\sqrt{E_b}}{\sqrt{2}\sigma} \; . \tag{16}$$

If this $Q$ factor were used in (2), the predicted BER would be much too worse (by approximately 3 dB) than the correct BER expression (15).

In order to predict the BER more accurately, we propose to estimate $Q$ in DPSK simulations by evaluating the variance of the field amplitude $|f_n|$ before the delay interferometer. We introduce an alternative "amplitude-$Q$", or $Q_A$, defined as

$$Q_A = \frac{\langle |f_n| \rangle}{\sigma_{|f_n|}} \approx \frac{\sqrt{E_b}}{\sigma} \; . \tag{17}$$

When we substitute $Q_A$ for $Q$ in (2), we find the BER result is consistent with (15) except for a less important factor in front of the exponential function. We note that $Q_A$ is larger than the result of (16) by a factor of $\sqrt{2}$ (or 3 dB).

### III. EFFECT OF NONLINEAR PHASE NOISE

So far we have restricted our discussion to the linear case. As the optical power increases, the above theory on BER estimation will eventually break down. One such nonlinear effect is the Gordon-Mollenauer phase noise which limits the performance of DPSK at high power levels [16]. In that case, $Q_A$ may be an overestimate of the performance and is no longer reliable.

When the BER is dominated by the such excessive phase noise, we find it helpful to evaluate $Q$ directly in the phase domain. Using polar coordinates $f_n = |f_n| e^{i\phi_n}$, we can rewrite (14) as

$$I_{\text{bal}} = |f_n f_{n-1}| \cos \Delta\phi_n . \quad (18)$$

Here, $\Delta\phi_n$ is the "differential phase" defined as

$$\Delta\phi_n = \phi_n - \phi_{n-1} , \quad (19)$$

which is distributed around either 0 (for "1") or $\pi$ (for "0"). The phase noise (including both the linear phase noise and the excessive phase noise due to nonlinearity) causes $\Delta\phi_n$ to deviate from its ideal value, and an error occurs if such deviation exceeds $\pi/2$. Suppose the standard deviation of $\Delta\phi_n$ (on both 0 and $\pi$) is $\sigma_{\Delta\phi}$, which can be easily computed in numerical simulations. We then introduce another alternative $Q$ definition, the "differential-phase-$Q$", or $Q_{\Delta\phi}$, as

$$Q_{\Delta\phi} = \frac{\pi}{2\sigma_{\Delta\phi}} \quad (20)$$

to estimate the BER when the transmission performance is limited by the nonlinear phase noise. This is based on the assumption that the fluctuation of $\Delta\phi_n$ has a Gaussian distribution. Although the exact distribution of the phase noise remains largely unknown at this point, the Gaussian approximation of the phase noise seems to capture the essence of the phase noise problem relatively well. For example, using this model we find that the nonlinear phase noise starts to impact the system performance when the total accumulated nonlinear phase shift of a WDM channel is on the order of 1 radian, as predicted by [16]. More studies on the nonlinear phase noise will be published elsewhere [17].

We note that in the low power regime $Q_{\Delta\phi}$ is larger than $Q_A$ by a factor of $\frac{\pi}{2\sqrt{2}}$ (or ~ 0.9 dB), and the BER is still dominated by $Q_A$. To take into account the contributions from both amplitude noise and phase noise as we vary the optical power levels in DPSK simulations, we choose the smaller one from $Q_A$ and $Q_{\Delta\phi}$ as the overall $Q$ for the BER estimation.

## IV. Conclusion

We have shown that in numerical simulations of DPSK, the $Q$ factor should be defined in the field domain, in contrast to the usual practice for OOK. Although we have discussed only an ideal case with matched filter detection, the principle can be generalized to more realistic systems with sub-optimum receivers. In addition, the excessive phase noise must be considered in the high power regime when the Kerr nonlinearity of the fiber plays a role. The exact nature of the noise distribution due to nonlinearity is yet to be explored.

## Acknowledgment

The authors gratefully acknowledge helpful discussions with R. E. Slusher, A. R. Chraplyvy, J. E. Mazo, J. Salz, G. Kramer, A. van Wijingaarden, L. F. Mollenauer, P. J. Winzer, A. H. Gnauck, R. Essiambre, S. Hunsche, T. I. Lakoba, and D. Fishman.